# Data-Driven Discovery of Population Balance Equations for the Particulate Sciences

SIMON ING XUN TIONG[a], FIRNAAZ AHAMED[b,c], YONG KUEN HO[a,d,e,1]

[a]Department of Chemical Engineering, School of Engineering, Monash University Malaysia, Jalan Lagoon Selatan, 47500 Bandar Sunway, Selangor, Malaysia

[b]School of Engineering, Faculty of Innovation and Technology, Taylor's University, 47500 Subang Jaya, Selangor, Malaysia

[c]Department of Biological Systems Engineering, University of Nebraska-Lincoln, Lincoln, Nebraska, USA

[d]Monash-Industry Plant Oils Research Laboratory, Monash University Malaysia, Jalan Lagoon Selatan, Bandar Sunway, 47500 Selangor, Malaysia

[e]Centre for Net-Zero Technology, Monash University Malaysia, Jalan Lagoon Selatan, Bandar Sunway, 47500 Selangor, Malaysia

[1]To whom correspondence should be addressed.

**Email**: ho.yongkuen@monash.edu,

**Abstract**

Understanding the behavior of particles in a dispersed phase system via population balances holds fundamental importance in studies of particulate sciences across various fields. Particle behavior, however, is sophisticated as a single particle can undergo internal property changes (e.g., size, cell age, and energy content) through various mechanisms. When confronted with an unknown distributed particulate system, discovering the underlying population balance equation (PBE) entails firstly learning the underlying particulate phenomena followed by the associated phenomenological laws that govern the kinetics and mechanisms of particle transformations in their local conditions. Conventional inverse problem approaches reveal the shape of phenomenological functions for predetermined forms of PBE (e.g., pure breakage/aggregation PBE, etc.). However, these methods can be limited in their ability to uncover the mechanisms which govern uncharacterized particulate systems from data. Leveraging the increasing abundance of data, we devise a data-driven framework based on sparse regression to learn PBEs as linear combinations of an extensive pool of candidate terms. Thus, this approach enables effective and accurate functional identification of PBEs without assuming the structure *a priori*, hence mitigating any potential loss of details, while minimizing model overfitting and providing a more interpretable representation of particulate systems. We showcase the proficiency of our approach across a wide spectrum of particulate systems, ranging from simple canonical pure breakage and pure aggregation systems to complex systems with multiple particulate processes. Our approach holds the potential to generalize the discovery of PBEs along with their phenomenological laws from data, thus facilitating wider adoption of population balances.

# 1. Introduction

A variety of diverse phenomena, such as the electrification of dust in storms, the morphological transformation of drug crystals, and the proliferation of biological cells, are all fundamentally driven by particles. Despite the contextual dissimilarity, the particles in each phenomenon share a common theme of being distributed in one or more properties, e.g., shapes and sizes of dusts and crystals, and age of biological cells. Of interest to scientists and engineers is to pursue the behavior of these collections of particles as their distributions evolve through time and space, to understand their dynamic behavior and implications. During the transient, individual particles may experience a loss of identity or assume a new identity by undergoing birth-death transformations (i.e., breakage and aggregation), and/or preserve it by experiencing convective process (i.e., growth) (1). Governing equations to describe these non-mutually exclusive evolutions are often expressed as integro-partial differential equations, or more commonly known as Population Balance Equations (PBEs).

Since the introduction of the first-ever PBE by Smoluchowski for a pure coagulation process in 1916 (2), the theoretical foundations of PBEs have matured considerably and can now be generalized to treat various growth, birth, and death processes of numerous entities. Implementations of PBEs have since been found in different fields, such as agricultural engineering (3-8), astrophysics and astronomy (9-11), and hematology (12-14). Although PBE is the fundamental equation that governs the temporal evolution of the particle distribution to preserve the number of distributed quantities, its success hinges on the fidelity of the phenomenological functions which describe the behavior of single particles in their local environment. Due to the lack of mechanistic knowledge, especially for relatively new or unknown systems, the development of these functions based on first principles is challenging. Although in principle the behavior of single particles should ideally be determined by direct experimental observation (15), in most cases it is still impractical to measure and track sufficient individual interactions to deduce the underlying phenomenological laws that govern the particle behaviors. Another approach to reveal the phenomenological functions would be to solve the so-called inverse problems, where the behavior of single particles is estimated from collective dynamic measurements (1, 15). Although existing inverse solution methodologies have the advantage of not being committed to any specific form of the model function, they either require the existence of specific traits of the data such as self-similar behavior (15-20) and characteristic behavior (21), or they result in highly ill-posed formulations (22). In the last decade, little advancement has been made to inverse problem methodologies, and existing methods are far from being capable of unveiling generalizable, and parsimonious functional forms of the phenomenological components to reveal interpretable systems, particularly for processes with multiple simultaneously occurring phenomena.

In light of the foregoing challenges which inhibit the widespread adoption of population balances to study new processes in particulate sciences, recent advances in data-driven discovery of dynamical systems with sparse regression present a radical potential to overcome these difficulties (23). Notably, recent advances in measurement technology, e.g., high-throughput micro-computed tomography (μCT), transmission electron microscopy (TEM), atomic force microscopy (AFM), and nanoparticle tracking analysis (NTA), have led to an increasing abundance of high-resolution data of particulate systems, rendering it exceptionally well-suited for the application of data-driven approaches using modern computing capacity. Unlike inverse solution methodologies which only attempt to uncover phenomenological functions in a pre-conceived form of PBE (e.g., pure breakage PBE), these data-driven approaches, which operate on the principle of parsimony, search through a high-dimensional solution space amongst a large library of candidate terms to select a minimal subset of terms that are quintessential to characterize systems of interest. Notably, the sparse identification of nonlinear dynamics (SINDy) and its variants (24-34), and partial differential equations functional identification of nonlinear dynamics (PDE-FIND) algorithms (35) have shown to be helpful in discovering ordinary and partial differential equations which govern various important physical systems. Due to the co-existence of birth-and-death processes during breakage or aggregation phenomena and the convective process which must satisfy a set of physical constraints (e.g., symmetry), directly adopting such approaches to learn PBEs from data is extremely non-trivial. Nevertheless, with extensive redesign guided by expert PBE knowledge, such approaches can form an integral part of a powerful PBE discovery tool that holds great promise for studies in the particulate sciences.

In this work, we utilize the concept of data-driven sparse model identification and introduce the **P**opulation **B**alance **E**quations Identification via **S**parsity **P**romoting **O**ptimization **T**echnique (PBE-SPOT) to identify PBEs from measured data of particulate systems. To prevent unrestricted identification of physically unrealizable PBEs, the PBE-SPOT searches across judiciously curated phenomenologically established libraries of candidate terms encompassing both integrals and partial derivatives and imposes a physics-informed selection procedure to arrive at a parsimonious and meaningful PBE. By achieving a generalizable and interpretable PBE, we can potentially discover the particulate processes (i.e., breakage, aggregation, and growth) which govern the given data without any *a priori* knowledge about the process. We demonstrate that the PBE-SPOT is applicable across an exhaustive number of particulate systems, from simple binary breakage system to systems where multiple processes occur simultaneously. As a key feature, PBE-SPOT employs a comprehensive screening strategy to allow various combinations of particulate processes to be included in the model selection procedure, rather than presuming the type of particulate processes in advance. Unlike the conventional approaches to identify PBEs, the PBE-SPOT marks a significant shift in the way we uncover PBEs from data, by being versatile, not reliant on any special traits in the data (e.g., self-similarity), and scalable to large-scale particulate systems.

## 2.0 Methods

## 2.1 One Dimensional Population Balance Equations for Describing Various Particulate Phenomena

The general 1D PBE for simultaneous growth, aggregation and breakage without spatial variation in the physical space can be written as follows:

$$\frac{\partial n(x,t)}{\partial t} + \frac{\partial}{\partial x}\left[R(x)n(x,t)\right] = \frac{1}{2}\int_0^x Q(x-y,y)n(x-y,t)n(y,t)dy - \int_0^\infty Q(x,y)n(x,t)n(y,t)dy \\ + \int_x^\infty \beta(x,y)\Gamma(y)n(y,t)dy - \Gamma(x)n(x,t) \quad (1)$$

Here, $n(x,t)$ is the continuous number density function where $t$ represents time, and $x$ refers to the internal coordinate of choice. In the context of this study, we use the terms 'internal coordinate' and 'particle size' interchangeably for ease of reference. The second term on the left-hand side (LHS) generally represents the growth of entities due to their convective motions along the internal coordinate space. The terms on the right-hand side (RHS), based on the order of their appearance, represent the birth and death functions for an aggregation process, and the birth and death functions for a breakage process. The phenomenological functions are embedded within the birth-and-death terms as well as the convective term, where $R(x)$ is the growth rate, $Q(x,y)$ is the aggregation kernel, $\beta(x,y)$ is the breakage stoichiometric function that describes the distribution of daughter particles resulting from breakage events and $\Gamma(x)$ is the breakage rate kernel.

For the sake of simplicity in the subsequent discussion, Eq. (1) can be re-written compactly in the following form:

$$\dot{n} = \Psi\left[B_{agg}(\bullet), D_{agg}(\bullet), B_{bkg}(\bullet), D_{bkg}(\bullet), G(\bullet)\right] \quad (2)$$

where $\dot{n} = \partial n(x,t)/\partial t$. Here, $\Psi$ represents a set function of linear and nonlinear terms that describe the behavior of the particle population. The set function $\Psi$ represents the exact solution characterizing the particulate system of interest, which may contain the aggregation terms $B_{agg}(\bullet) = \frac{1}{2}\int_0^x (\bullet) n(x-y,t) n(y,t) dy$ and $D_{agg}(\bullet) = \int_0^\infty (\bullet) n(x,t) n(y,t) dy$, the breakage terms $B_{bkg}(\bullet) = \int_x^\infty (\bullet) n(y,t) dy$ and $D_{bkg}(\bullet) = (\bullet) n(x,t)$, the growth term $G(\bullet) = \partial[(\bullet) n(x,t)]/\partial x$, or any of their combinations thereof.

## 2.2 Population Balance Equation Identification via Sparsity Promoting Optimization Technique (PBE-SPOT)

The PBE-SPOT seeks to discover meaningful and interpretable PBEs along with the phenomenological functions from transient data of particle distributions. Unlike the conventional partial differential equation discovery which may be subject to indeterminate functional forms, the PBE-SPOT leverages the structure of PBEs with established integrals (i.e., for the birth-and-death processes due to aggregation and breakage) and partial derivatives with respect to the internal coordinate (i.e., for the convective process due to growth) to identify the PBE which underlie a given set of data. With this knowledge of the equation architecture, we can include a broad spectrum of candidate functions $(\bullet)$ into the integrals and partial derivatives that are potentially contained within $\Psi$. Our primary postulate is that a sparse linear combination of these candidate functions can fairly approximate the phenomenological functions (or kernels), and thereby accurately depict the particle behavior over time. Moreover, based on the principle of Occam's razor (36), the optimal set function $\Psi$ is anticipated to be the one with the fewest number of terms possible while not compromising model accuracy.

Given a set of experimental number density data with *j* discretized points in the internal coordinate space, and collected for *k* temporal points, it becomes possible to set up a linear system of equations as follows to determine $\Psi$:

$$\dot{\mathbf{n}} = \Omega\left[\Theta_{agg}, \Theta_{bkg}, \Theta_{G}\right]\xi \tag{3}$$

Here, $\dot{\mathbf{n}} = \left[\dot{n}(x_1, t_1) \quad \dot{n}(x_2, t_1) \quad \cdots \quad \dot{n}(x_{j-1}, t_k) \quad \dot{n}(x_j, t_k)\right]^T$ is a column vector that contains the temporal partial derivatives that are numerically approximated from the transient number density distribution. The embedding of a wide range of candidate functions into the integrals and partial derivatives creates the sub-libraries for aggregation, breakage, and growth terms, denoted as $\Theta_{agg}$, $\Theta_{bkg}$ and $\Theta_{G}$, respectively. As an example, the breakage sub-library $\Theta_{bkg}$ which contains the birth-and-death candidate terms, is as shown below:

$$\Theta_{bkg} = \left[\begin{array}{cccccccc} | & | & | & & | & | & | & \\ B_{bkg}(\mathbf{1}) & B_{bkg}(\mathbf{y}) & B_{bkg}(\mathbf{y}^2) & \cdots & D_{bkg}(\mathbf{1}) & D_{bkg}(\mathbf{x}) & D_{bkg}(\mathbf{x}^2) & \cdots \\ | & | & | & & | & | & | & \end{array}\right]\Big\} j \times k \tag{4}$$

For ease of representation, we employ notations such as $\mathbf{y}^2$ and $\mathbf{x}^2$ to denote the column vectors which contain the values of power functions $y^2$ and $x^2$, respectively. Certainly, the functions embedded in the candidate terms are not limited to simple power functions, for example, more complicated forms of functions such as exponential function and piecewise function may also be embedded in the sub-libraries, but their inclusion should be guided by meticulous analysis of the available data. To provide a clearer illustration of the notations, example columns of $\Theta_{bkg}$ evaluated for the entire data set are shown below for the birth-and-death terms with quadratic candidate terms:

$$B_{bkg}(\mathbf{y}^2) = \left[ \int_{x_1}^{\infty} y^2 n(y,t_1) dy \quad \int_{x_2}^{\infty} y^2 n(y,t_1) dy \quad \cdots \quad \int_{x_{j-1}}^{\infty} y^2 n(y,t_k) dy \quad \int_{x_j}^{\infty} y^2 n(y,t_k) dy \right]^T \quad (5)$$

$$D_{bkg}(\mathbf{x}^2) = \left[ x_1^2 n(x_1,t_1) \quad x_2^2 n(x_2,t_1) \quad \cdots \quad x_{j-1}^2 n(x_{j-1},t_k) \quad x_j^2 n(x_j,t_k) \right]^T \quad (6)$$

Note that the candidate function $(\cdot)$ in $B_{bkg}(\cdot) = \int_x^{\infty} (\cdot) n(y,t) dy$ represents the product of the stoichiometric function and breakage rate kernel. All integrals and partial derivatives can be approximated numerically via methods like the trapezoidal rule and finite difference formulae, respectively. These sub-libraries of choice can then be concatenated adjacently to produce a master library $\Omega$ which is then curated to eliminate linear dependencies (Supplementary Material, Section S1.2). The full linear combination of all the columns of the curated master library $\Omega$ is clearly inapt for describing the particle behavior because the exact solution might carry extraneous terms and result in a highly overfitted model. Such an overfitted population balance model fails to generalize and make meaningful predictions on the transient of particle distribution. Thus, we intend to select only a sparse subset of columns in the curated $\Omega$ to arrive at a parsimonious PBE, which aptly describes the relevant phenomena occurring in the system of interest. The column vector $\xi$ in Eq. (3) represents a sparse vector where its non-zero entries determine the active columns in the curated $\Omega$. The sparse selection of appropriate PBEs using PBE-SPOT is discussed in detail in the following section. For more details on setting up PBE-SPOT for PBE identification, we refer readers to Section S1.0 of the Supplementary Materials.

## 2.3 Sparse Equation Learning and Evaluation for Choosing the Optimal Realizable Population Balance Equations (SELECT-PBE)

To identify physically realizable PBEs from the curated $\Omega$, we employ the highly efficient sparse regression algorithm, i.e., Sequential Thresholded Least Square (STLS) (24) to generate a pool of sparse solutions using a wide range of sparsity index λ for all possible non empty combinations of sub-libraries designed in this work, i.e., pure growth, pure breakage, pure aggregation, and all their combinations thereof. This approach was motivated by the inconsistent performance of the STLS when used directly on the full library encompassing all potential physics (cf. Supplementary Material, Section S1.1). Given the efficiency of STLS, using it to generate a reasonably large (but finite) pool of sparse solutions is generally computationally inexpensive. The optimal model is then selected from this large pool of solutions based on the following cost function:

$$\xi = \underset{\xi'}{\operatorname{argmin}} \ \lambda_1 \|\boldsymbol{\Omega}\xi' - \dot{\mathbf{n}}\|_2^2 + \lambda_2 \|\xi'\|_0 + \lambda_3 \Phi(\xi') \tag{7}$$

where $\lambda_i$ ($i = 1,2,3$) is the individual weights in the cost function and $\boldsymbol{\Omega}$ is, based on the preceding discussion, the curated master library. We call this procedure the Sparse Equation Learning and Evaluation for Choosing the Optimal Realizable Population Balance Equations (SELECT-PBE). Here, the first two terms on the RHS represent the $l_0$-regularization, and the third term is introduced to penalize the solutions with invalid PBE structures. The motivation for such an approach is twofold, the first is to search for potential equations in smaller solution spaces of different combinations of sub-libraries such that the most relevant phenomena (or phenomenon) which describe the data can be determined. Second, within a specific phenomenon, the STLS being inherently unconstrained could potentially arrive at PBEs with unrealizable physics, i.e., the existence of birth terms without corresponding deaths and vice versa. Such formulations contradict the law of mass conservation and number preservation in particulate processes. The SELECT-PBE hinders this by imposing additional penalty scores to unrealizable PBEs via the third term of Eq. (7). The procedure here can be perceived as a reframed $l_0$-regularization problem, as the optimal solution with the lowest penalty score will be selected from the predetermined pool of solutions. We show in detail that adding the third term is essential to improve model identification in the Supplementary Material, Section S1.1. It is important to highlight that the SELECT-PBE does not only work exclusively with STLS, and alternative sparse regression methods can be equally employed.

**2.4 Formulating Identified Population Balance Equations from Physical Interpretation**

Upon successfully solving for the sparse vector in Eq. (7), a PBE can be formulated as follows:

$$\dot{n} = \boldsymbol{\omega}^T \boldsymbol{\xi} \tag{8}$$

Here, $\boldsymbol{\omega}$ is a symbolic column vector which contains the mathematical description of all the columns in the curated $\boldsymbol{\Omega}$. The entries of the symbolic vector are merely the mathematical representation of the columns such as $B_{bkg}(x)$, $D_{bkg}(x)$, $B_{agg}(x)$, which elucidate both the particulate process and the embedded candidate function.

It should be noted that the master library $\boldsymbol{\Omega}$ might be rank deficient and potentially introduce bias to the outcome. To address such an issue, we devise an algorithm to eliminate the columns that demonstrate linear dependency with other columns in the library while preserving information about the removed column/term. If the entries in $\boldsymbol{\omega}$ involve those that exhibit linear dependency, e.g., the elements of column 2 are twice the elements of column 5 in the un-curated $\boldsymbol{\Omega}$ whereby column 5 is deliberately eliminated due to its linear dependency, then the corresponding element in $\boldsymbol{\omega}$ is the mathematical expression represented by 'column 2 or two multiplied by column 5'. When such situations arise, this information contained in $\boldsymbol{\omega}$ allows one to make informed decisions on the final form of PBE based on meaningful interpretation of the overall equation structure. Details on eliminating column dependencies in the library can be found in the Supplementary Material, Section S1.2.

## 3. Results

### 3.1 Discovery of Various Population Balance Equations

We employed the PBE-SPOT to identify a total of 16 different particulate systems, ranging from single process (e.g., pure aggregation, pure breakage, and pure growth), to competing parallel processes (e.g., simultaneous growth and aggregation), and opposing processes (e.g., simultaneous breakage and aggregation). The data used here was generated from analytical solutions (37-39), and in instances where they were not available (i.e., simultaneous breakage and aggregation), we resorted to the Fixed Pivot Technique (40). In addition, an exponential initial condition, i.e., $n_0(x) = \exp(x)$, was employed for all cases to ensure a fair comparison. In this work, the functional forms of terms in the sub-libraries used for PBE identification were chosen based on the physics of the respective phenomena: power functions and reciprocal power functions up to an order of 3 are incorporated in $\Theta_{bkg}$ and $\Theta_G$, whereas power functions and product functions (i.e., $x^2y$, $xy^2$) up to a total degree of 3 are included in $\Theta_{agg}$. The foregoing configuration gives a total of 41 terms in the uncurated master library $\Omega$. We opted for lower-order functions here since higher-order functions often lack generalizability, leading to overfitting.

Table 1 shows the summary of the results of PBE-SPOT for various particulate systems. Here, we computed the average error of the coefficients for the identified PBEs from both clean and noisy data. We also evaluated the robustness of PBE-SPOT in handling noisy data by adding white noise to data at specified magnitudes. To smoothen the noisy data, we employed the

Savitzky-Golay filter. Moreover, we also used a polynomial interpolation strategy (35) to further evaluate the derivatives of filtered noisy data. In contrast, no additional treatments were applied to evaluate the integrals of noisy data as the errors are somewhat offset by the alternating upward and downward trend. The specified noise level (i.e., 0.25%, 0.5%, 0.75%, and 1%) in Table 1 refers to the maximum noise level at which the PBEs remain identifiable (at least >10% success rate, which is the percentage of successful identification of all terms over 100 runs), beyond which identification fails. Further details on the impact of noise and data sampling on the performance of PBE-SPOT are provided in Section S2 in the Supplementary Material. From the results, except for Cases (j) and (l), all particulate systems are identified with good accuracy from clean data, i.e., less than 5% average error. Out of the 16 cases examined on clean data, only Case (l) involving simultaneous F$x+y$ breakage and sum aggregation is not fully identifiable, as the algorithm identifies every term correctly with good accuracy but consistently misses the aggregation death term. In a similarly challenging situation of multi-process Case (j), all the terms are identified correctly but with larger average error, which we discuss further in Section 3.4. Nevertheless, being able to identify most of the terms accurately in Case (l) or all of the terms but with inaccurate coefficients in Case (j) allows one to gain useful insights on these complex multi-phenomena systems which may be used to further interrogate and refine the model. In this work, when subjected to noisy data, in general systems with pure breakage exhibit higher tolerance, while the systems driven by aggregation and multiple processes are more susceptible to noise – the latter as shown by Cases (n), (o) and (p) being hardly identifiable when the data is corrupted with noise. It is noteworthy that rank deficiency also occurs with noisy data, which further affirms the need for the column elimination algorithm to arrive at precise and interpretable PBE.

In particulate sciences, the number density distribution often evolves across multiple orders of magnitude with time, making it crucial to provide data that carries significant dynamics for model identification using PBE-SPOT. Due to variations in the underlying dynamics of the particulate systems, the sampling of data for accurate identification unavoidably varies on a case-by-case basis (more details in Supplementary Material, Section S2). In the context of population balances, it should be noted that number densities of extremely small magnitudes, e.g., $10^{-8}$, must not be disregarded despite appearing insignificant. For instance, the number density of large particles is often small but is critical in describing size-expanding systems.

Similarly, the range and sampling frequency in the temporal domain are equally important, as it is crucial that the data captures the significant and rapid dynamics in the system. Nevertheless, our results show that in all test cases, PBE-SPOT is generally more sensitive to the choice of the time range and therefore identifies the correct solutions when provided with an appropriate range, with a finer temporal grid being employed only when necessary. This implies that in practice, through model testing and validation, one could deploy the PBE-SPOT to screen through different time ranges and granularity of data (and possibly across a range of optimization weights $\lambda_i$) to arrive at the correct PBE. We show the appropriate domain along with the SELECT-PBE working parameters for each case in Section S1.3 of the Supplementary Material.

In the subsequent sections, we delve deeper into the identification of PBEs for several phenomena, exploring special cases and demonstrating the robustness of our approach by examining extreme cases.

Table 1: Summary of the PBE-SPOT results for a variety of particulate systems with clean and noisy data. The specified noise level in each scenario represents the maximum limit beyond which correct terms cannot be identified. Errors are reported in the form of average errors of the coefficients. The operators $B_{agg}(\cdot)$, $D_{agg}(\cdot)$, $B_{bkg}(\cdot)$, $D_{bkg}(\cdot)$, and $G(\cdot)$ are defined in Section 2.1. Further details on the setup for each case can be found in Section S1.3 of the Supplementary Material. Effect of noise and data sampling are explored in Section S2.

| System | PBE and Error w/o noise, w/ noise (noise level) | System | PBE and Error w/o noise, w/ noise (noise level) |
|---|---|---|---|
| (a) Constant Aggregation 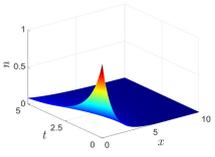 | $\dot{n} = B_{agg}(1) - D_{agg}(1)$<br><br>0.18%, 0.20% ± 0.20% (1%) | (b) Sum Aggregation 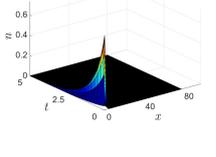 | $\dot{n} = B_{agg}(x) - D_{agg}(x+y)$<br><br>3.72%, 2.30% ± 2.05% (1%) |
| (c) Product Aggregation 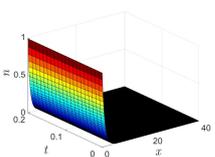 | $\dot{n} = B_{agg}(xy - y^2)$<br>$- D_{agg}(xy)$<br><br>1.53%, 2.04% ± 2.74% (0.25%) | (d) Fx+y Breakage† 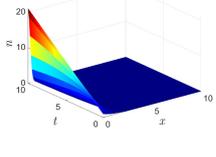 | $\dot{n} = 2B_{bkg}(y) - D_{bkg}(x^2)$<br><br>0.57%, 0.54% ± 0.40% (1%) |
| (e) F1 Breakage† 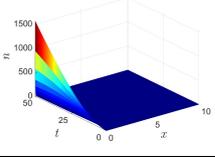 | $\dot{n} = 2B_{bkg}(1) - D_{bkg}(x)$<br><br>1.46%, 1.46% ± 0.75% (1%) | (f) Constant Growth 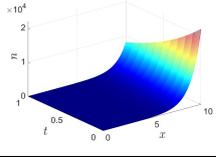 | $\dot{n} = -G(1)$<br><br>0.05%, 3.25% ± 0.75% (1%) |
| (g) Linear Growth 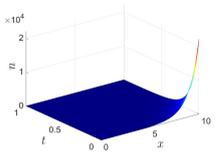 | $\dot{n} = -G(x)$<br><br>0.12%, 0.30% ± 0.42% (0.5%)# | (h) Linear Growth and Constant Aggregation 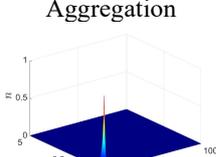 | $\dot{n} = -G(x) + B_{agg}(10)$<br>$- D_{agg}(10)$<br><br>0.66%, 77.18% ± 1.15% (0.25%) |

| System | PBE and Error w/o noise, w/ noise (noise level) | System | PBE and Error w/o noise, w/ noise (noise level) |
|---|---|---|---|
| (i) Linear Growth and Sum Aggregation 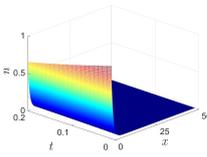 | $\dot{n} = -G(x) + B_{agg}(x) - D_{agg}(x+y)$<br><br>0.46%, 1.76% ± 1.70% (0.25%) | (j) Constant Growth and Aggregation 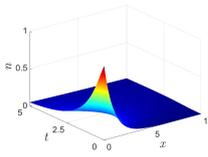 | $\dot{n} = -G(1) + B_{agg}(1) - D_{agg}(1)$<br><br>27.51%, ‡ |
| (k) Fx+y Breakage and Constant Aggregation 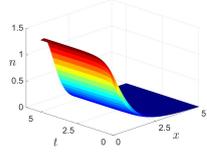 | $\dot{n} = 2B_{bkg}(y) - D_{bkg}(x^2) + B_{agg}(1) - D_{agg}(1)$<br><br>3.45%, 3.44% ± 0.13% (0.5%) | (l) Fx+y Breakage and Sum Aggregation* 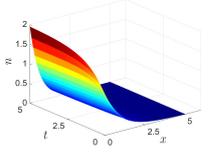 | $\dot{n} = 2B_{bkg}(y) - D_{bkg}(x^2) + B_{agg}(x) - D_{agg}(x+y)$<br><br>– |
| (m) Fx+y Breakage and Product Aggregation 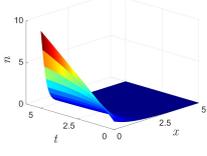 | $\dot{n} = 2B_{bkg}(y) - D_{bkg}(x^2) + B_{agg}(xy - y^2) - D_{agg}(xy)$<br><br>3.26%, 3.48% ± 2.77% (0.5%) | (n) F1 Breakage and Constant Aggregation§ 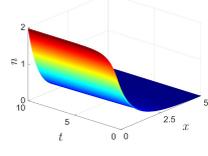 | $\dot{n} = 2B_{bkg}(1) - D_{bkg}(x) + B_{agg}(1) - D_{agg}(1)$<br><br>1.14%, – |
| (o) F1 Breakage and Sum Aggregation§ 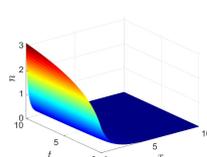 | $\dot{n} = 2B_{bkg}(1) - D_{bkg}(x) + B_{agg}(x) - D_{agg}(x+y)$<br><br>3.20%, – | (p) F1 Breakage and Product Aggregation§ 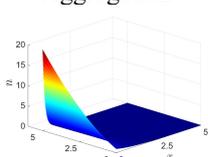 | $\dot{n} = 2B_{bkg}(1) - D_{bkg}(x) + B_{agg}(xy - y^2) - D_{agg}(xy)$<br><br>4.42%, – |

† Cases (d) and (e) are referred to as 'F1' and 'Fx+y', respectively, due to their kernels being inherently intertwined and can be represented as single intrinsic rates, i.e., $F(x,y) = 1$ and $F(x,y) = x+y$ (38).
# The error and standard deviation are reported for the case where 20% of the full dataset is utilized.
‡ Case (j) cannot be identified when noisy data was used, primarily due to the omission of $G(1)$. Only $B_{agg}(1)$ and $D_{agg}(1)$ are identified.
* Case (l) cannot be fully identified using PBE-SPOT, as STLS consistently misses $D_{agg}(y)$.
§ Cases (n), (o), and (p) exhibit high susceptibility to noise.

## 3.2 Identifying Aggregation with Sum Kernel from Multiple Potential Solutions

Aggregation is a process where two distinct particles coalesce and give rise to the birth of a larger, new particle. It plays a crucial role in various applications, such as crystallization and precipitation of pharmaceutical substances (41), flocculation (42), granulation (43), emulsion polymerization (44). Due to its prevalence in various processes, unraveling the aggregation kernel from data has been a persistent problem since the 1980s. Various mathematical properties, i.e., function homogeneity and asymptotic behavior of the similarity distribution (15, 17, 20), have been utilized to deduce the aggregation kernel from data in the past. Such mathematical properties are often based on simplified assumptions and special traits of the system, which might be challenging to derive in more complicated scenarios. To showcase the capability of PBE-SPOT in identifying aggregation PBEs, we first employ it to study a very canonical example, in which the particles merge at a rate that is equivalent to the sum of their sizes, i.e., $Q(x, y) = x + y$.

$$\frac{\partial n(x,t)}{\partial t} = \frac{1}{2}\int_0^x xn(x-y,t)n(y,t)dy - \int_0^\infty (x+y)n(x,y)n(y,t)dy \tag{9}$$

Fig. 1 shows the schematic workflow for the identification of the sum aggregation PBE using the PBE-SPOT. Using number density data ranging between $x \in [0.01, 20.01]$ with an interval of 0.1, and $t \in [0,1]$ with an interval of 0.1, the first step involves constructing the libraries of different combinations of phenomena. Our algorithm reveals the presence of linearly dependent columns in the un-curated master library following the procedure detailed in Section S1.2 of the Supplementary Material. Of particular significance is that $B_{agg}(x)$ or $2B_{agg}(y)$ are identified as mutually exclusive potential birth terms by the SELECT-PBE from the pool of predetermined solutions. From Eq, (1), the aggregation frequencies for the birth term and death term are given as

$Q(x-y, y)$ and $Q(x, y)$ respectively. The aggregation kernel of $Q(x, y)$ indicates the rate at which particles of sizes $x$ and $y$ merge to form new particles of size $x+y$, while $Q(x-y, y)$ describes the rate of formation of particles of size $x$ due to the aggregation of sizes $x - y$ and $y$. As both the birth and death terms must share identical mechanism of aggregation, upon further scrutiny, one can conclude that $B_{agg}(x)$ is a more appropriate birth term as it is congruent with the sum kernel identified in the death term, i.e., $Q(x, y) = x + y$ implies that $Q(x-y, y) = x$. Thus, when dealing with an aggregation system, it is important to ensure that the aggregation frequencies embedded in the birth-and-death terms share the same mathematical implications, as they must jointly and uniformly govern the process. If the identified aggregation frequencies yield different implications, both of them must be considered in tandem to arrive at a more accurate function that describes the aggregation behavior of particles.

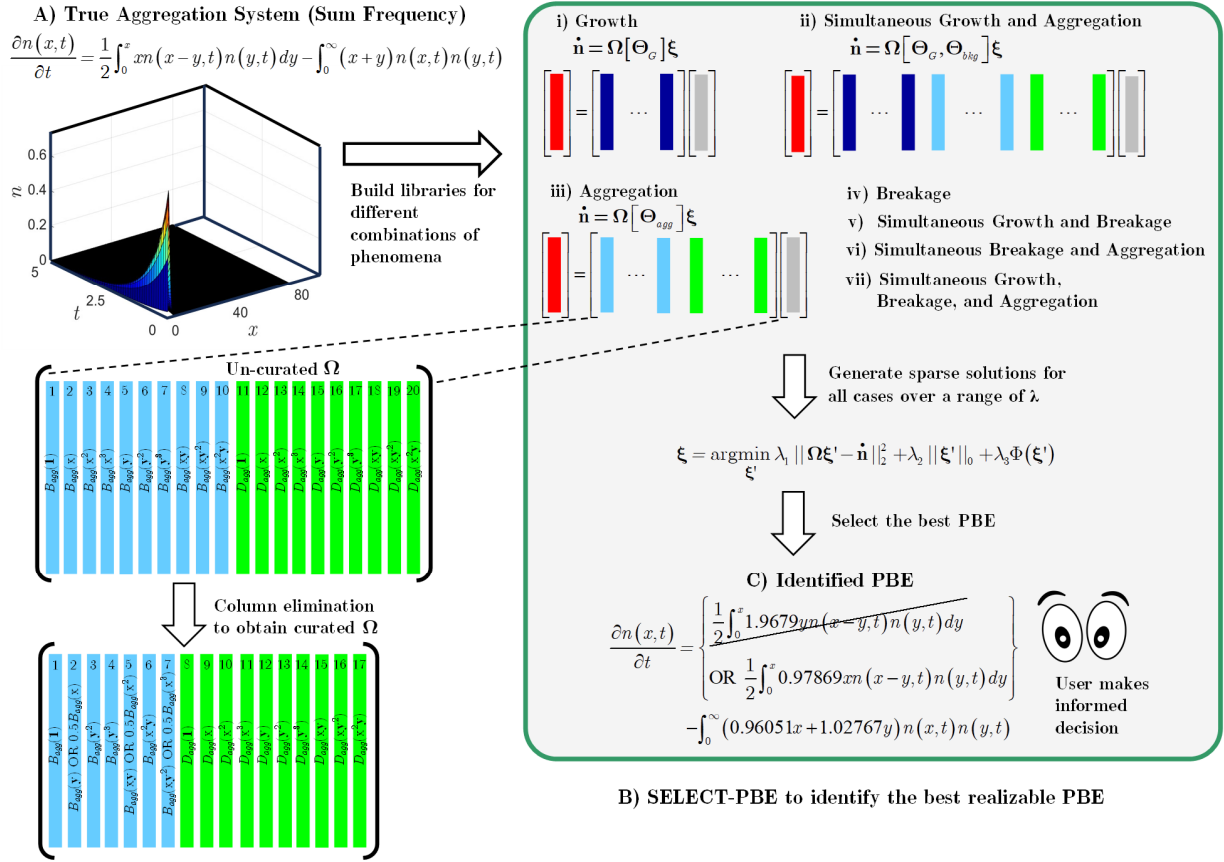

Fig. 1 Schematic workflow of the PBE-SPOT, illustrated using a pure aggregation system with sum kernel. The first step is to collect the time series number density distribution data and numerically approximate the time derivatives $\dot{\mathbf{n}}$. Then, un-curated master libraries $\Omega$ with different combinations of particulate phenomena are constructed. Column elimination algorithm is subsequently applied to produce less rank-deficient curated master libraries $\Omega$. Next, sparse solutions are generated across a range of sparsity index for all the curated master libraries. The SELECT-PBE then selects the optimal model amongst the pool of solutions with the lowest penalty score. In this illustrative example, $2B_{agg}(y)$ OR $B_{agg}(x)$ is identified as mutually exclusive potential birth term of the optimal model. Finally, the $B_{agg}(x)$ term is selected due to its more realizable description.

### 3.3 Decoupling the Stoichiometric Kernel and Breakage Rate Kernel For Binary Breakage

Next, we consider another canonical PBE, i.e., random binary breakage. Random binary breakage is a process in which a single parent particle randomly fragments into two daughter particles under the influence of external forces, e.g., in depolymerization (45), granulation, and milling. In the context of pure breakage, the stoichiometric kernel and the breakage rate kernel dictate the particle behavior. The conventional approach relies on self-similarity behavior of the system to extract these underlying functions (16), which again may not be applicable when dealing with a more complicated system. The PBE which describes the evolution of particle distribution under random binary breakage (i.e., $\beta(x,y) = 2/y$, where the underlying daughter distribution function follows a uniform distribution) with a linear breakage rate (i.e., $\Gamma(x) = x$) is given as:

$$\frac{\partial n(x,t)}{\partial t} = 2\int_x^\infty n(y,t)dy - xn(x,t) \tag{10}$$

where the coefficient of '2' in the birth term denotes the average number of daughter particles produced from the breakage events, implying a binary breakage process. In this identification exercise, we omit the physical coefficient '2' in formulating $\Theta_{bkg}$ to allow PBE-SPOT to derive the information from the data, rather than imposing it *a priori*. With number density data ranging between $x \in [0.01, 5.01]$ with an interval of 0.1, and $t \in [0,5]$ with an interval of 0.1, PBE-SPOT accurately retrieves the breakage PBE, along with the coefficient of 2 in the birth term.

However, identifying the stoichiometric kernel in the case of breakage is more intricate, as it is intertwined through multiplication with the breakage rate kernel in the birth term. Fortunately, the breakage rate kernel can be directly deduced from the death term, i.e., $\Gamma(x) = x$ and it will assist in deducing the stoichiometric kernel. Since, $2B_{bkg}(1)$ is identified as the birth term, mathematically it implies the relationship $\beta(x,y)\Gamma(y)=2$, where it can be deduced that $\beta(x,y)=2/y$. However, in scenarios where the stoichiometric kernel follows a more complicated marginal probability function, its retrieval is non-trivial. In such scenarios, longer chains of polynomials may be identified as the birth and death kernels when only simple power functions are used in the candidate library. This naturally leads to a rational function when attempting to deduce the stoichiometric kernel. While rational function might be empirically applicable within the specified internal coordinate domain, it's instrumental to seek a more interpretable form of the stoichiometric kernel, which is potentially achievable via fitting the resulting rational function to established marginal probability functions that satisfy the mass/number preservation implied by breakage process.

**3.4 PBE-SPOT Identifies Systems with Multiple Processes**

It is not uncommon for particulate systems to exhibit multiple processes simultaneously. The case where particles grow in size while simultaneously merging with other particles presents a challenge for model identification. Although growth and aggregation are distinct processes, the transient data does not easily distinguish between them, as both result in size enlargement. Deducing both the growth rate and aggregation kernel within a single framework is a complex task that existing inverse solution approaches have yet to accomplish. One exception is the method developed by Mahoney, Doyle III and Ramkrishna (21) for simultaneous growth and aggregation systems, where the growth law can be extracted only on the condition that the aggregation dynamics are known.

Here, we employ the PBE-SPOT to identify a challenging system where both growth rate and aggregation kernel cause particles to expand in size at a constant rate. The similarity in the rate constants makes it difficult to distinguish between growth and aggregation despite their distinct moment-related implications, e.g., growth conserves particle number whereas aggregation decreases it. The PBE that describes the process is given as:

$$\frac{\partial n(x,t)}{\partial t} + \frac{\partial}{\partial x}\left[n(x,t)\right] = \frac{1}{2}\int_0^x n(x-y,t)n(y,t)dy - \int_0^\infty n(x,y)n(y,t)dy \qquad (11)$$

With number density data ranging between $x \in [0.01, 10.01]$ with an interval of 0.01, and $t \in [0,5]$ with an interval of 0.1, our PBE-SPOT identifies the correct term but with an observable margin of error (~ 25%, Table 1) in the coefficients. Moreover, when a higher noise level is imposed, only the aggregation terms with constant frequencies are identified but with unsatisfactory coefficients. This occurs because the dynamics of the system closely resemble that of a pure constant aggregation system, and the difficulty in distinguishing them is exacerbated by noise. The ability of PBE-SPOT to overcome this challenge is highly valuable, providing users with insights into the underlying particulate phenomena and enabling further model testing and refinement. In cases where growth and aggregation functions exhibit distinct trends (cases (h) and (i) in Table 1), they are identified with a much lower margin of error. The results highlight the capability of PBE-SPOT in handling systems with concurrent processes, encountering challenges only when managing parallel processes with similar rates, yet still effectively identifying part of the correct terms for further evaluation.

## 4.0 Discussion

We have established a data-driven method called PBE-SPOT for discovering PBEs for various particulate phenomena. Our approach fundamentally differs from existing inverse problem approaches by directly uncovering a wide variety of PBEs from data, rather than merely identifying phenomenological functions within a pre-determined form of PBE. Our approach considers two key aspects for effective PBE identification: (1) employment of sparse regression technique to distill the potential PBEs from various combinations of sub-libraries for distinct particulate processes, and (2) selection of the optimal and physically realizable PBE using reframed $l_0$-regularization. Using an exhaustive number of test cases from various combinations of particulate processes, our approach has the potential to address the long-standing challenge to obtain generalizable, interpretable, and parsimonious phenomenological functions for PBEs from dynamically evolving distribution data of particulate systems. In particular, our PBE-SPOT accurately recovers the PBEs along with their underlying phenomenological functions for various combinations of particulate phenomena within a unified framework. This contrasts with conventional approaches, which require varied treatment for different particulate phenomena and fall short in handling systems with multiple phenomena.

Despite particulate distribution data being capable of evolving across many orders of magnitude with time and containing a potentially wide range of underlying temporal frequencies (e.g., for size-dependent breakage process), we have demonstrated that with appropriate coverage of the significant data dynamics, the underlying PBEs can be identified using PBE-SPOT. Unfortunately, generalizing the sampling procedure (i.e., the range and granularity of data) for extracting the significant data is highly complex due to its dependency on the convolution of the initial condition of particle population, the type of particulate processes involved, the underlying kernels governing the system, etc. To the best of our knowledge, generalization of the sampling procedure has yet to be explicitly addressed in the sparse system identification and PBE inverse problem literature, which warrants a more systematic and reliable method to extract important data. Nevertheless, expert empirical knowledge of the system may be used to select different cuts of data to be supplied to the PBE-SPOT to systematically search for the optimal and physically realizable model. The PBE-SPOT is also reasonably robust when implemented on noisy systems with a single process in this work. More importantly, being able to identify the correct terms (albeit inaccurate coefficients) for many challenging cases with multiple processes at noisy conditions is a remarkable starting point towards generalizable, interpretable, and parsimonious population balance model discovery. To improve the robustness of the algorithm to handle noise, one may also consider adopting the weak form formulation of the PBEs, or incorporate new strategies to mitigate the impact of noise.

Existing sparse regression PDE discovery methods pose challenges in selecting appropriate measurement coordinates from unfamiliar high-dimensional space with unrestricted possible forms of candidate library functions (24). On the contrary, the PBE-SPOT search procedure enables an intuitive selection of the appropriate internal coordinate from particulate data. By leveraging the known structure of PBEs, our approach effectively reduces the number of potential library candidate terms, thereby preventing prohibitive computational expenses. As the PBEs identified using SELECT-PBE are always physically realizable (provided that the pool of solutions contains at least one realizable PBE), the phenomenological laws for the identified phenomenon can be readily distilled as linear combinations of the relevant candidate terms. More sophisticated functions such as the exponential or logarithmic functions (or their products) can also be included in the library, but this should be introduced with caution in the presence of other simple power functions as the former can often be expressed as an infinite series of the latter. There is no simple solution to arrive at the appropriate sparsifying function basis, for which physical intuition may be combined with other advanced feature extraction tools to design the library functions. Although the interdependence among the candidate functions could impede the discovery of a sparse basis, insights from failure to identify a sparse representation could be used to revise the library functions. Beyond the context of the form of PBEs discussed in this study, the framework of the PBE-SPOT can also be readily extended to spatially homogenous multi-dimensional population balances, for which the number density evolution of two or more internal coordinates can be tracked with time. Cases that require coupling with equations for the continuous phase (e.g., with the Navier-Stokes equation) will require further deliberation but are extremely powerful to study multi-phase fluid-particle interacting systems, such as the aerosol drug delivery, multiphase microfluidics for cancer detection, the evolution of the interstellar medium.

The growing abundance of data in the last decade has made way for data science to emerge as an important tool for scientific discovery. We leverage this emerging opportunity and develop the PBE-SPOT to accelerate the discovery of dynamical equations for the particulate sciences. As with any data-driven tool, PBE-SPOT is far from being a cookbook solution to all problems in the particulate sciences, however, having this tool at our disposal provides a principled approach to study data of the particulate sciences. A particularly impactful example is to deploy the PBE-SPOT to continuously track the evolution of the form of PBEs (and the corresponding shift in the underlying phenomena) as experimental conditions vary over a wide range, by which the insights obtained may be used to accelerate efforts to attain a more universal approximation. In this regard, having a PBE discovery algorithm that is less prone to overfitting and promotes generalizability and interpretability is of significant value. We foresee that the ideas presented here will serve as the foundation for future developments, and this will encourage wider adoption of PBEs in new and emerging fields of the particulate sciences, such as nanotechnology, synthetic biology, environmental science, etc.

## Acknowledgements

We would like to acknowledge the financial support in the form of Fundamental Research Grant Scheme (FRGS/1/2020/TK0/MUSM/03/1) from the Ministry of Higher Education Malaysia (MOHE), of which Yong Kuen Ho is the principal investigator.

**Supplementary Material**

**S1.0 Identification of Population Balance Equations using Sparse Regression**

Given a set of experimental number density data with $j$ discretized points in the internal coordinate space, collected for $k$ time points, it becomes possible to set up a linear system of equations as follows to determine $\Psi$ of Eq. (2) in the main text:

$$\dot{\mathbf{n}} = \Omega \left[ \Theta_{agg}, \Theta_{bkg}, \Theta_G \right] \xi \tag{S1}$$

where $\xi$ is the sparse vector where its non-zero entries indicate the active columns in the master library $\Omega$. To obtain $\dot{\mathbf{n}}$, we first collect the particle number density distribution at different time points (**N**), numerically compute the time derivatives at different time points ($\dot{\mathbf{N}}$) on account of the LHS of Eq. (1) in the main text using any numerical differentiation approach, and reshape $\dot{\mathbf{N}}$ into a column vector $\dot{\mathbf{n}}$, as shown below:

$$\mathbf{N} = \begin{bmatrix} n(x_1,t_1) & n(x_1,t_2) & \cdots & n(x_1,t_k) \\ n(x_2,t_1) & n(x_2,t_2) & \cdots & n(x_2,t_k) \\ \vdots & \vdots & \ddots & \vdots \\ n(x_j,t_1) & n(x_j,t_2) & \cdots & n(x_j,t_k) \end{bmatrix} \tag{S2a}$$

$$\dot{\mathbf{N}} = \begin{bmatrix} \dot{n}(x_1,t_1) & \dot{n}(x_1,t_2) & \cdots & \dot{n}(x_1,t_k) \\ \dot{n}(x_2,t_1) & \dot{n}(x_2,t_2) & \cdots & \dot{n}(x_2,t_k) \\ \vdots & \vdots & \ddots & \vdots \\ \dot{n}(x_j,t_1) & \dot{n}(x_j,t_2) & \cdots & \dot{n}(x_j,t_k) \end{bmatrix} \tag{S2b}$$

$$\dot{\mathbf{n}} = \begin{bmatrix} \dot{n}(x_1,t_1) \\ \dot{n}(x_2,t_1) \\ \dot{n}(x_2,t_1) \\ \vdots \\ \dot{n}(x_{j-1},t_k) \\ \dot{n}(x_j,t_k) \end{bmatrix} \quad (S2c)$$

In the above, the dimension of the master library $\boldsymbol{\Omega}$ is $p \times q$, where $p = j \times k$ and $q$ is the total number of candidate functions in the master library. In this work, we assume that the phenomenological functions for particulate processes can be approximated to a reasonable degree via polynomial functions, rational functions or any type of functions that can be formed from the linear combination of basic functions such as 1, $x$, $x^2$, $1/x$, $y$, $y^2$, $1/y$, $xy$, $x^2y^3$. According to the convention adopted in this work, $x$ and $y$ are dummy variables that represent the internal coordinate (e.g., size, cell age, volume) with respect to the parent and daughter particles. As an example of how a sub-library may be built, consider the aggregation sub-library $\boldsymbol{\Theta}_{agg}$ which may be built by embedding basic functions into $B_{agg}(\bullet)$ and $D_{agg}(\bullet)$ to obtain the following:

$$\boldsymbol{\Theta}_{agg} = \begin{bmatrix} | & | & | & & | & | & & | & | & \\ B_{agg}(\mathbf{1}) & B_{agg}(\mathbf{x}) & B_{agg}(\mathbf{x}^2) & \cdots & B_{agg}(\mathbf{y}) & B_{agg}(\mathbf{y}^2) & \cdots & B_{agg}(\mathbf{xy}) & B_{agg}(\mathbf{x}^2\mathbf{y}) & \cdots \\ | & | & | & & | & | & & | & | & \\ | & | & | & & | & | & & | & | & \\ D_{agg}(\mathbf{1}) & D_{agg}(\mathbf{x}) & D_{agg}(\mathbf{x}^2) & \cdots & D_{agg}(\mathbf{y}) & D_{agg}(\mathbf{y}^2) & \cdots & D_{agg}(\mathbf{xy}) & D_{agg}(\mathbf{x}^2\mathbf{y}) & \cdots \\ | & | & | & & | & | & & | & | & \end{bmatrix} \Big\} j \times k$$

(S3)

Example columns of Eq. (S3) evaluated for the entire data set are shown below for selected birth-and-death terms:

$$B_{agg}\left(\mathbf{x}^2\right) = \begin{bmatrix} \frac{1}{2}\int_0^{x_1}\left(x_1^2\right)n(x_1-y,t_1)n(y,t_1)dy \\ \frac{1}{2}\int_0^{x_2}\left(x_2^2\right)n(x_2-y,t_1)n(y,t_1)dy \\ \vdots \\ \frac{1}{2}\int_0^{x_{j-1}}\left(x_{j-1}^2\right)n(x_{j-1}-y,t_k)n(y,t_k)dy \\ \frac{1}{2}\int_0^{x_j}\left(x_j^2\right)n(x_j-y,t_k)n(y,t_k)dy \end{bmatrix} \Big\} j \times k \quad (S4)$$

$$D_{agg}\left(\mathbf{xy}\right) = \begin{bmatrix} \int_0^{\infty}(x_1 y)n(x_1,t_1)n(y,t_1)dy \\ \int_0^{\infty}(x_2 y)n(x_2,t_1)n(y,t_1)dy \\ \vdots \\ \int_0^{\infty}(x_{j-1} y)n(x_{j-1},t_k)n(y,t_k)dy \\ \int_0^{\infty}(x_j y)n(x_j,t_k)n(y,t_k)dy \end{bmatrix} \Big\} j \times k \quad (S5)$$

All integrals are numerically approximated using the trapezoidal rule or Simpson's rule, whereas the derivatives for convective processes are estimated numerically via finite difference. Upon constructing the master library $\mathbf{\Omega}$ by including the desired sub-libraries, we further overcome the possibility of encountering a rank-deficient $\mathbf{\Omega}$, which could potentially distort the outcome, by devising an algorithm to eliminate the columns that demonstrate linear dependency with other columns in the library (Section S1.2). With $\dot{\mathbf{n}}$ and the curated $\mathbf{\Omega}$, based on the aggregation example above, if the true underlying process is aggregation with product kernel, and with birth-and-death kernels of $Q(x-y,y) = xy - y^2$ and $Q(x,y) = xy$, respectively, solving Eq. (S1) for the sparse vector $\xi$ will trigger the columns represented by $B_{agg}(\mathbf{xy})$, $B_{agg}(\mathbf{y}^2)$, and $D_{agg}(\mathbf{xy})$ in the master library by assigning them with non-zero coefficients.

Upon successfully solving for the sparse vector, a PBE can be formulated as follows:

$$\dot{n} = \boldsymbol{\omega}^T \boldsymbol{\xi} \tag{S6}$$

Here, $\boldsymbol{\omega}$ is a symbolic column vector which contains the mathematical description of the active columns in the curated $\boldsymbol{\Omega}$. The entries of the symbolic vector elucidate both the particulate process and the candidate function embedded within the specific process. For instance, the entry $B_{agg}(x^2)$ represents the birth process due to aggregation embedded with a quadratic function. The depository of mathematical description as a form of symbolic vector becomes useful when mathematical manipulations are required to attain a more interpretable solution (Section S1.2).

**S1.1 Sparse Equation Learning and Evaluation for Choosing the Optimal Realizable PBEs (SELECT-PBE)**

In theory, the sparse solution of Eq. (S1) can be obtained by solving the $l_0$-regularization problem described by the first two terms on the RHS of the following cost function (with $\lambda_1$ equals unity):

$$\boldsymbol{\xi} = \underset{\boldsymbol{\xi}'}{\mathrm{argmin}} \ \lambda_1 \| \boldsymbol{\Omega}\boldsymbol{\xi}' - \dot{\mathbf{n}} \|_2^2 + \lambda_2 \| \boldsymbol{\xi}' \|_0 + \lambda_3 \Phi(\boldsymbol{\xi}') \tag{S7}$$

The second term on the RHS is the $l_0$-norm, which is defined as the number of non-zero entries in $\boldsymbol{\xi}$. However, the $l_0$-regularization is widely recognized as NP-hard because it is mathematically intractable and comes with the inherent issue of combinatorial explosion when dealing with high-dimensional systems. Moreover, we found that solving the convex relaxation problems of the $l_0$-regularization, i.e., LASSO and Ridge regression, is inefficient to provide a sparse basis for the overdetermined library $\mathbf{\Omega}$ in our study.

To efficiently compute the sparse solution, we initially employed the Sequential Thresholded Least-Squares (STLS). The STLS algorithm is a computationally economical and simple method which produces the least-squares solution by zeroing the entries that are smaller than the sparsity index $\lambda$ after each least squares iteration until the remaining entries are all greater than $\lambda$. However, the STLS tends to select the incorrect terms and identify a solution with invalid PBE formulation when all the particulate processes ($\mathbf{\Theta}_{agg}$, $\mathbf{\Theta}_{bkg}$, $\mathbf{\Theta}_G$) are incorporated in the library. Table 1 shows the identification of different particulate systems by employing different variants of curated $\mathbf{\Omega}$, where only a few PBEs can be identified successfully when all particulate processes are included. Therefore, the likelihood of success for STLS to identify the correct PBE decreases when irrelevant particulate processes are included in the library. For instance, the STLS can only identify most of the pure aggregation PBEs when the working library contains solely $\mathbf{\Theta}_{agg}$. It is important to note that merely increasing the extent and sampling rate of data will not convert the failed cases into successful ones, which further affirms that the root cause of failed identification lies in the combined influence of the extensive library and the sparse regression employed. Moreover, without any physical constraints, the STLS can freely identify any columns in the curated master library $\mathbf{\Omega}$ as being dominant which leads to the possibility of selecting the birth term independently of the death term and vice versa. In the context of population balances, the birth-and-death processes are mathematically intertwined and therefore inseparable in order to preserve the underlying number balance of entities.

Table S1: The identification of each particulate system under different variants of curated $\Omega$. STLS could not identify the case of simultaneous Fx+y breakage and sum aggregation (clean data).

| | $\Omega[\Theta_G]$ | $\Omega[\Theta_{bkg}]$ | $\Omega[\Theta_{agg}]$ | $\Omega\begin{bmatrix}\Theta_G,\\ \Theta_{bkg}\end{bmatrix}$ | $\Omega\begin{bmatrix}\Theta_G,\\ \Theta_{agg}\end{bmatrix}$ | $\Omega\begin{bmatrix}\Theta_{bkg},\\ \Theta_{agg}\end{bmatrix}$ | $\Omega\begin{bmatrix}\Theta_G,\\ \Theta_{bkg},\\ \Theta_{agg}\end{bmatrix}$ |
|---|---|---|---|---|---|---|---|
| Constant Aggregation | - | - | ✓ | - | ✗ | ✓ | ✗ |
| Sum Aggregation | - | - | ✓ | - | ✓ | ✗ | ✗ |
| Product Aggregation | - | - | ✓ | - | ✓ | ✗ | ✗ |
| F+xy Breakage | - | ✓ | - | ✓ | - | ✓ | ✓ |
| F1 Breakage | - | ✓ | - | ✓ | - | ✓ | ✓ |
| Constant Growth | ✓ | - | - | ✗ | ✓ | - | ✗ |
| Linear Growth | ✓ | - | - | ✓ | ✓ | - | ✗ |
| Linear Growth and Constant Aggregation | - | - | - | - | ✓ | - | ✗ |
| Linear Growth and Sum Aggregation | - | - | - | - | ✓ | - | ✓ |
| Constant Growth and Aggregation | - | - | - | - | ✓ | - | ✗ |
| Fx+y Breakage and Constant Aggregation | - | - | - | - | - | ✓ | ✗ |
| Fx+y Breakage and Sum Aggregation | - | - | - | - | - | ✗ | ✗ |
| Fx+y Breakage and Product Aggregation | - | - | - | - | - | ✓ | ✗ |
| F1 Breakage and Constant Aggregation | - | - | - | - | - | ✓ | ✗ |
| F1 Breakage and Sum Aggregation | - | - | - | - | - | ✓ | ✗ |
| F1 Breakage and Product Aggregation | - | - | - | - | - | ✓ | ✗ |

The foregoing shortcoming of the STLS poses challenges when dealing with a new system with unknown particle behaviors. Without any *a priori* knowledge, the occurrence of breakage, aggregation, or growth, either individually or in various combinations, is unknown in a new system. To deal with this issue, we retain the STLS as the method for sparse regression but overcome its weakness for discovering PBEs by enacting a comprehensive screening strategy referred to as the Sparse Equation Learning and Evaluation for Choosing the Optimal Realizable PBE (SELECT-PBE) framework. The SELECT-PBE, being the workhorse for PBE-SPOT, first constructs different variants of $\boldsymbol{\Omega}$ by including combinations of different sub-libraries with different particulate processes, i.e., a) pure growth $\boldsymbol{\Omega}[\boldsymbol{\Theta}_G]$, b) pure breakage $\boldsymbol{\Omega}[\boldsymbol{\Theta}_{bkg}]$, c) pure aggregation $\boldsymbol{\Omega}[\boldsymbol{\Theta}_{agg}]$, d) simultaneous growth and breakage $\boldsymbol{\Omega}[\boldsymbol{\Theta}_G, \boldsymbol{\Theta}_{bkg}]$, e) simultaneous growth and aggregation $\boldsymbol{\Omega}[\boldsymbol{\Theta}_G, \boldsymbol{\Theta}_{agg}]$, f) simultaneous breakage and aggregation $\boldsymbol{\Omega}[\boldsymbol{\Theta}_{bkg}, \boldsymbol{\Theta}_{agg}]$, and g) simultaneous growth, breakage and aggregation $\boldsymbol{\Omega}[\boldsymbol{\Theta}_G, \boldsymbol{\Theta}_{bkg}, \boldsymbol{\Theta}_{agg}]$. Then, the STLS is used to systematically compute the sparse solutions for every variant of curated $\boldsymbol{\Omega}$ under a specified range of $\lambda$. This approach yields a pool of sparse solutions, of which the next task is to select the one that best represents the system of interest by evaluating all the solutions using the cost function shown in Eq. (S7). In addition to the classical $l_0$-regularization, the cost function in Eq. (S7) contains an additional third term which penalizes the solutions that are physically unrealizable.

While evaluating a $l_0$-regularization cost function over a finite number of sparse solutions is computationally tractable, the solution with the lowest penalty score may not always exhibit a sound structure due to STLS being unconstrained. The sole inclusion of only either the birth or death term defies the law of conservation of mass and number balance in particulate systems. To prevent this, ideally, it would be desirable to constrain the search by incorporating known properties of PBEs such as the conservation of the first moment in a non-reacting system for breakage and/or aggregation processes. However, doing so loses the generality of the framework, as the conservation of the first moment does not apply to growth processes. Moreover, even for a pure breakage/aggregation process, the range of applicable data for successful identification does not necessarily reflect the mass conservation property, and including the full range that does might skew the outcome, thereby hindering the inclusion of moment-based properties in the cost function. In search for the simplest solution, we observe that solely imposing penalties for physically unrealizable PBEs is sufficient to induce a significant improvement in the success of PBE identification. The last term of Eq. (S7), dubbed as the SELECT-PBE regularizer, does this by checking the physical realizability of each PBE in the large pool of solutions and increases the penalty scores for those that violate physical laws.

To exemplify the significance and intricacies of the SELECT-PBE regularizer, we employ the case of aggregation with summation kernel as an illustrative example. In the absence of the SELECT-PBE regularizer, with equal weights being assigned to both $\lambda_1$ and $\lambda_2$, a physically unrealizable PBE, i.e., $-1.3 D_{agg}(y)$, is identified. In such a scenario, one might posit that increasing $\lambda_1$ for the least square error would be beneficial, but it results in another physically unrealizable PBE, i.e., $-B_{bkg}(1) - D_{agg}(x) - D_{agg}(y)$. Here, further adjusting both $\lambda_1$ and $\lambda_2$ is futile as the physically unrealizable PBE has a slightly lower least square error and it holds the same number of terms as the correct model, i.e., $B_{agg}(x) - D_{agg}(x) - D_{agg}(y)$. Therefore, unless an additional term is present in the cost function, the physically realizable and true model remains unattainable. The SELECT-PBE regularizer operates as follows: when breakage or aggregation birth term is detected without their respective death terms and vice versa in the solution, it will introduce an additional penalty to the solution, thereby making it less favorable or impossible to be selected. The value of the penalty score of the regularizer is user-defined, where we set it as unity in this work. Table S2 shows the identifiability of the various PBEs studied in this work with/without the SELECT-PBE regularizer. The cases of pure aggregation consistently cannot be identified without the presence of the regularizer. While it may appear useful only in cases involving aggregation, given the potential abundance of unrealizable PBEs in the solution pool as compared to the canonical examples in this work, the utility of the SELECT-PBE regularizer can extend to real-world problems by filtering them out.

Table S2: The identifiability of various PBEs studied in this work with/without the SELECT-PBE regularizer, $\Phi(\xi')$, where '✗' indicates failure to identify the PBE despite having tuned the weights of the least-square error and the $l_0$-norm, and '✓' denotes successful identification.

| | without $\Phi(\xi')$ | with $\Phi(\xi')$ |
|---|---|---|
| Constant Aggregation | ✗ | ✓ |
| Sum Aggregation | ✗ | ✓ |
| Product Aggregation | ✗ | ✓ |
| Fx+y Breakage | ✓ | ✓ |
| F1 Breakage | ✓ | ✓ |
| Constant Growth | ✓ | ✓ |
| Linear Growth | ✗ | ✓ |
| Linear Growth and Constant Aggregation | ✓ | ✓ |
| Linear Growth and Sum Aggregation | ✓ | ✓ |
| Constant Growth and Aggregation | ✓ | ✓ |
| Fx+y Breakage and Constant Aggregation | ✓ | ✓ |
| Fx+y Breakage and Sum Aggregation | ✗ | ✗ |
| Fx+y Breakage and Product Aggregation | ✓ | ✓ |
| F1 Breakage and Constant Aggregation | ✓ | ✓ |
| F1 Breakage and Sum Aggregation | ✓ | ✓ |
| F1 Breakage and Product Aggregation | ✓ | ✓ |

**S1.2 Elimination of Linearly Dependent Columns**

It is not uncommon to encounter a rank-deficient master library $\Omega$ for system identification using sparse regression. This may distort the outcome of PBE identification using PBE-SPOT in the absence of appropriate intervention. For integro-partial differential equations, we observe that linear dependence can even occur among a group of two or more columns. To address this issue, we first detect the linearly independent columns (46) in the library. Then, we extract the linearly dependent groups of columns from the library and compute their null basis to determine their relationships. As an illustrative example, in the case of aggregation with sum kernel, given the columns which are sequentially represented by $B_{agg}(\mathbf{1})$, $B_{agg}(\mathbf{x})$ and $B_{agg}(\mathbf{y})$, computing the null basis using MATLAB for this group of columns reveals that linear dependence manifests in the form of $0.01 B_{agg}(\mathbf{1}) + B_{agg}(\mathbf{x}) = 2 B_{agg}(\mathbf{y})$. We thus regard the $B_{agg}(\mathbf{1})$ column as the weakly correlated column, , and establish $B_{agg}(\mathbf{x})$ and $B_{agg}(\mathbf{y})$ as being strongly dependent. In such a situation, there is a chance that STLS identifies the column represented by $B_{agg}(\mathbf{y})$ and completely disregards $B_{agg}(\mathbf{x})$. This poses an issue where $B_{agg}(\mathbf{x})$ is the more interpretable term but was disregarded due to underlying selectivity of STLS.

Our strategy is to remove one of the dependent columns in each group to form the curated master library $\Omega$ while registering the relationships in the symbolic vector $\omega$ so that users can later select the more interpretable term based on physical knowledge. In the foregoing example where $B_{agg}(\mathbf{x})$ is removed from the master library, the corresponding $B_{agg}(\mathbf{y})$ column in the symbolic vector $\omega$ will register the mathematical description of '$B_{agg}(y)$ OR $0.5 B_{agg}(x)$'. As expected, the PBE-SPOT identifies $2B_{agg}(y)$ as the aggregation birth term. This implies that $B_{agg}(x)$ term is also identified along. As the identified death term is $D_{agg}(x) + D_{agg}(y)$ which implies $Q(x,y) = x + y$, for a user familiar with population balances, it is easy to perceive that $B_{agg}(x)$ is clearly more interpretable than $2B_{agg}(y)$ due to $Q(x-y, y) = x$. Fig. S1 shows the workflow of the column elimination algorithms discussed here.

In essence, for linearly dependent pairs, either member of the pair can be removed whereas in the case where three columns are linearly dependent, we design the algorithm to remove one of the strongly correlated columns. The rank-deficient problem occurs for some of the cases in this study, where it could cause the incorrect PBEs and phenomenological kernels to be identified in the absence of remedial measures. Examples of this problem occurring are the cases where sum and product aggregation frequencies are governing the process. The column elimination approach presented in this section discerns the underlying relationship between all the candidate terms and is crucial to the success of the PBE-SPOT. Lastly, the devised algorithm is also useful for other unprecedented physical systems where the exact equation representation is still unknown. In this case, a diversity of terms (e.g., higher order derivatives, integrals, basic functions) may be needed, the algorithm can hence deduce the inherent mathematical connections between the candidate terms with respect to the system and facilitate the identification of the true and interpretable system.

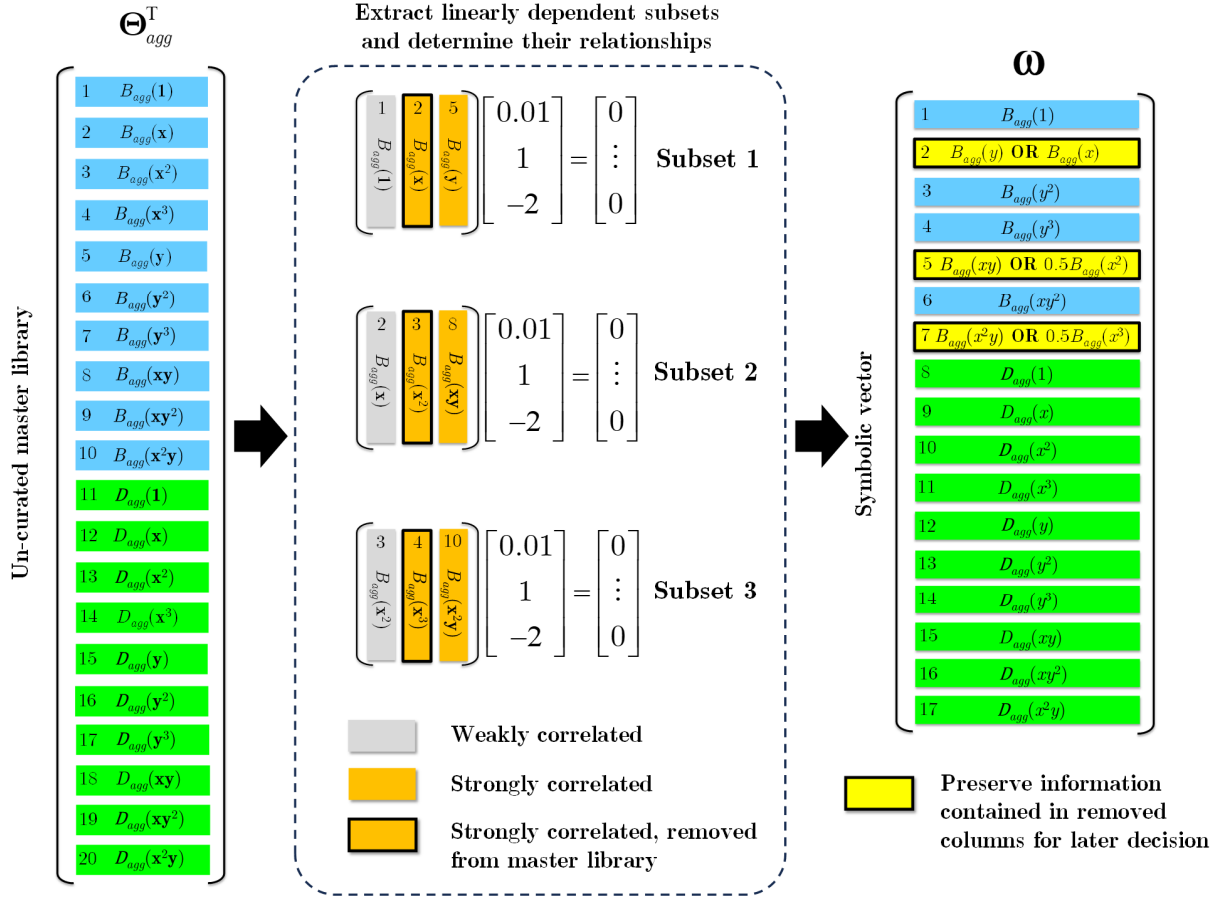

Fig. S1 Illustrative workflow of the column elimination algorithm, exemplified using the aggregation library ($\Omega\left[\Theta_{agg}\right]$) for the case of aggregation with sum kernel. The algorithm commences by extracting all the subsets of linearly dependent columns from the library while determining their relationships through computing their null basis. In this example, columns 1, 2 and 3 are considered the weakly correlated column in their respective set. The algorithm is designed to eliminate one of the strongly correlated columns in the subset. Thus, the algorithm eliminates columns 2, 3 and 4 from their respective subset (and consequently from the master library) while preserving their information in the symbolic vector $\omega$.

**S1.3 Specification of Domains and SELECT-PBE Parameters**

Table S3 shows the internal coordinate domain, temporal domain and the parameters of SELECT-PBE for the successful identification of each case study in this work. It is worth mentioning that the range of data covered here does not reflect the moment-based properties, e.g., the conservation of first moment over time, the increase in zeroth moment due to breakage. The reflection of these properties requires a very fine and extended internal coordinate domain, which from our experience does not necessarily lead to improved outcome of sparse regression at extended time durations. In this work, we employ a fine internal coordinate domain (i.e., with an of interval = 0.01) only when necessary; otherwise, a coarse grid is preferred due to its lower computational cost. It can be observed that the cases which involve only size-increasing processes, e.g., growth and aggregation, often require a more extended grid to capture its important dynamics as they consistently cause number density changes at the larger size region. Conversely, fine temporal grid is consistently applied in this work and it allows the system's dynamics to be captured adequately. However, further work is needed to carefully investigate the impact of the temporal domain and explore the strategies for its reduction, leading to a more efficient solution.

Table S3: The specifications of the internal coordinate domain and temporal domain, along with parameters of SELECT-PBE across all case studies in this work.

| | **Internal coordinate domain** | **Temporal domain** | **SELECT-PBE Parameters** $[\lambda_1, \lambda_2, \lambda_3]$ |
|---|---|---|---|
| Constant Aggregation | $x \in [0.01, 10.01], \Delta x = 0.01$ | $t \in [0, 5], \Delta t = 0.1$ | $[1,1,1]$ |
| Sum Aggregation | $x \in [0.01, 20.01], \Delta x = 0.1$ | $t \in [0, 1], \Delta t = 0.1$ | $[1,1,1]$ |
| Product Aggregation | $x \in [0.01, 25.1], \Delta x = 0.1$ | $t \in [0, 0.2], \Delta t = 0.01$ | $[1,1,1]$ |
| Fx+y Breakage | $x \in [0.1, 10.1], \Delta x = 0.1$ | $t \in [0, 10], \Delta t = 0.01$ | $[1,1,1]$ |
| F1 Breakage | $x \in [0.1, 5], \Delta x = 0.1$ | $t \in [0, 5], \Delta t = 0.1$ | $[1,1,1]$ |
| Constant Growth | $x \in [0.01, 10], \Delta x = 0.01$ | $t \in [0, 1], \Delta t = 0.1$ | $[0.01, 1, 1]$ |
| Linear Growth | $x \in [0.01, 10], \Delta x = 0.01$ | $t \in [0, 1], \Delta t = 0.01$ | $[0.01, 1, 1]$ |
| Linear Growth and Constant Aggregation | $x \in [0.01, 100.01], \Delta x = 0.1$ | $t \in [0, 5], \Delta t = 0.01$ | $[1,1,1]$ |
| Linear Growth and Sum Aggregation | $x \in [0.01, 50], \Delta x = 0.01$ | $t \in [0.01, 0.2], \Delta t = 0.01$ | $[2,1,1]$ |
| Constant Growth and Aggregation | $x \in [0.01, 10.01], \Delta x = 0.01$ | $t \in [0.01, 5], \Delta t = 0.01$ | $[1,2,1]$ |
| Fx+y Breakage and Constant Aggregation | $x \in [0.01, 5.01], \Delta x = 0.1$ | $t \in [0, 5], \Delta t = 0.01$ | $[2,1,1]$ |
| Fx+y Breakage and Sum Aggregation | $x \in [0.01, 5.01], \Delta x = 0.1$ | $t \in [0, 5], \Delta t = 0.01$ | $[1,2,1]$ |
| Fx+y Breakage and Product Aggregation | $x \in [0.01, 5.01], \Delta x = 0.1$ | $t \in [0, 5], \Delta t = 0.01$ | $[1,1,1]$ |
| F1 Breakage and Constant Aggregation | $x \in [0.01, 5], \Delta x = 0.01$ | $t \in [0, 10], \Delta t = 0.01$ | $[1,1,1]$ |
| F1 Breakage and Sum Aggregation | $x \in [0.01, 10], \Delta x = 0.01$ | $t \in [0, 10], \Delta t = 0.01$ | $[1,1,1]$ |
| F1 Breakage and Product Aggregation | $x \in [0.01, 5], \Delta x = 0.01$ | $t \in [0, 5], \Delta t = 0.01$ | $[1,1,1]$ |

**S2.0 Noise Level and Data Sampling Size Sensitivity of PBE-SPOT**

In this section, we examined the performance of the PBE-SPOT across different noise levels and data sampling sizes for all cases presented in this work, except cases (l), (n)-(p). The noise level ranges from 0.25% to 1%, whereas the percentage of the data drawn from the full data indicated in the Table S3 spans between 20% to 100%. We evaluated each noise level and data sampling size with 100 random samples from the full data following a uniform distribution. Here, we define the success rate of PBE-SPOT as the number of correct identification of all the terms. The success rate of identifying the correct PBE terms for all cases is demonstrated in the following figures (Figs. S2 – S6).

PBE-SPOT shows the most success with the noisy breakage systems in this work, but its performance deteriorates as the amount of data used is reduced. It also works well against the constant aggregation case even when the amount of data is scarce due to its simplistic kernel. The sum aggregation case is quite robust against noise but not identifiable when the data used is insufficient. Product aggregation cases are very susceptible to noise. Moreover, it is not surprising that the constant growth and aggregation PBE is very challenging to identify when the data is corrupted, i.e., the growth term cannot be extracted and instead only the aggregation terms are identified. Furthermore, we would like to highlight most of the cases with opposing dynamics (i.e., simultaneous breakage and aggregation) are particularly challenging to analyze when the data is corrupted and limited, especially the cases where F1 breakage and aggregation coupled (i.e., cases (n)-(p)). While PBE-SPOT is robust in handling various kinds of uncorrupted particulate systems, instances where the data is corrupted and limited require more careful consideration. It is worth mentioning that the incorrect identification results may provide insights into the nature of the true results, i.e., some parts of the incorrect solution could be correct. Future work should focus on mitigating the impact of noise on the identification process, adopt and develop new strategies for dealing with limited data. Both of them are critical for ensuring PBE-SPOT's robustness and applicability across various conditions.

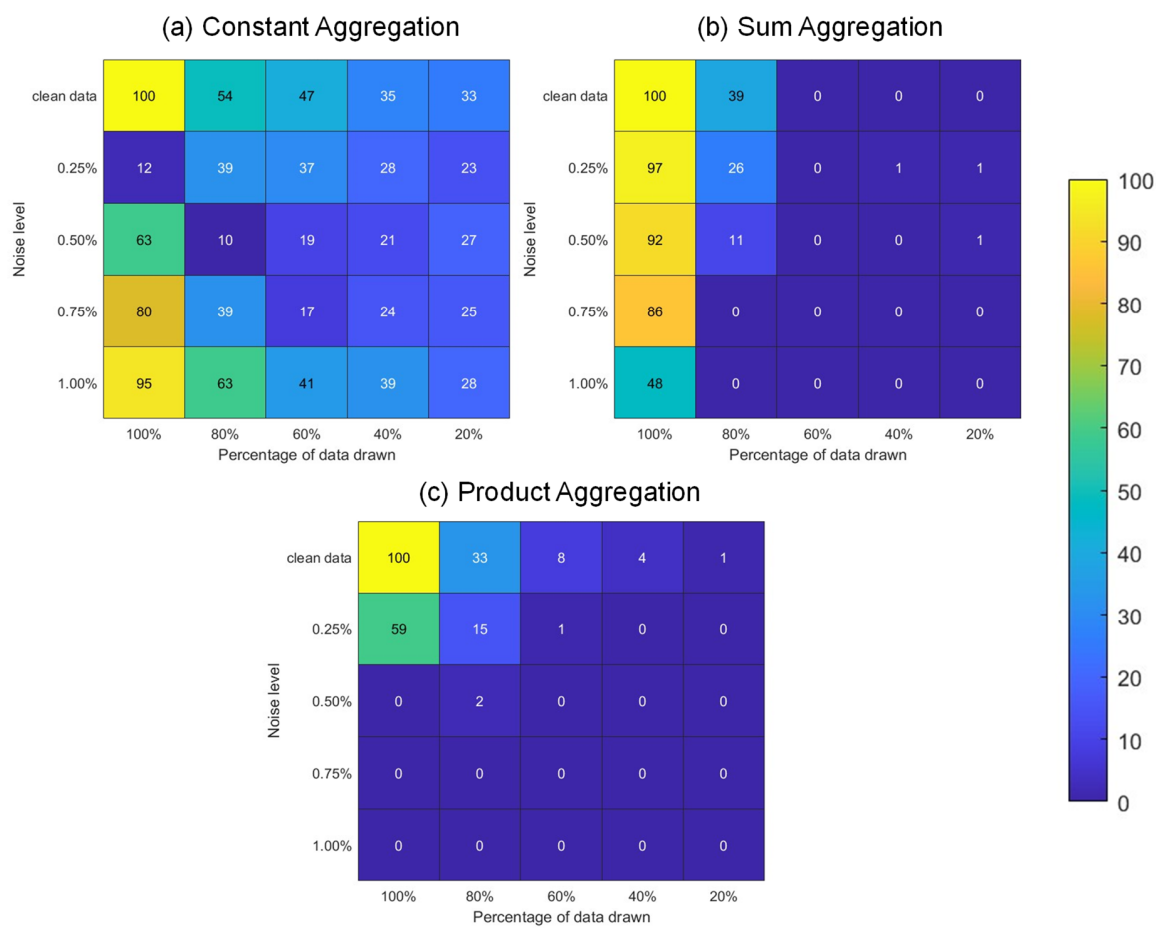

Fig. S2 The success rate (correct identification of PBE terms) of PBE-SPOT on pure aggregation cases: (a) constant aggregation, (b) sum aggregation and (c) product aggregation, across varying noise levels and amount of data. The noise levels tested are 0% (clean data), 0.25%, 0.50%, 0.75% and 1%. The amount of data used are 20%, 40%, 60%, 80% and 100% (data domain as indicated in Section S1.3). For each combination of noise level and data percentage, 100 random samples were tested.

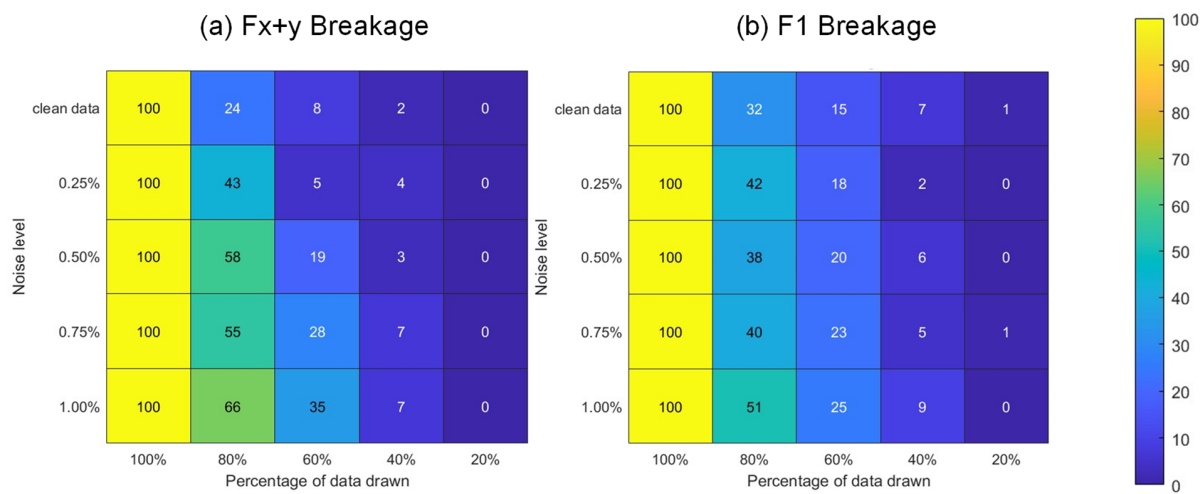

Fig. S3 The success rate (correct identification of PBE terms) of PBE-SPOT on pure breakage cases: (a) Fx+y breakage and (b) F1 breakage, across varying noise levels and amount of data. The noise levels tested are 0% (clean data), 0.25%, 0.50%, 0.75% and 1%. The amount of data used are 20%, 40%, 60%, 80% and 100% (data domain as indicated in Section S1.3). For each combination of noise level and data percentage, 100 random samples were tested.

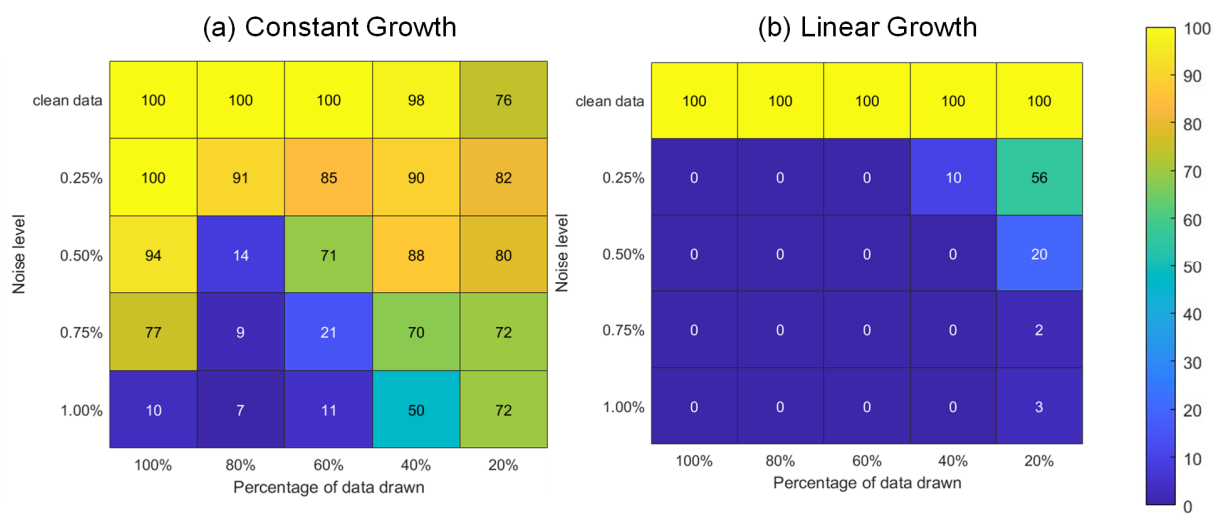

Fig. S4 The success rate (correct identification of PBE terms) of PBE-SPOT on pure growth cases: (a) constant growth and (b) linear growth, across varying noise levels and amount of data. The noise levels tested are 0% (clean data), 0.25%, 0.50%, 0.75% and 1%. The amount of data used are 20%, 40%, 60%, 80% and 100% (data domain as indicated in Section S1.3). For each combination of noise level and data percentage, 100 random samples were tested.

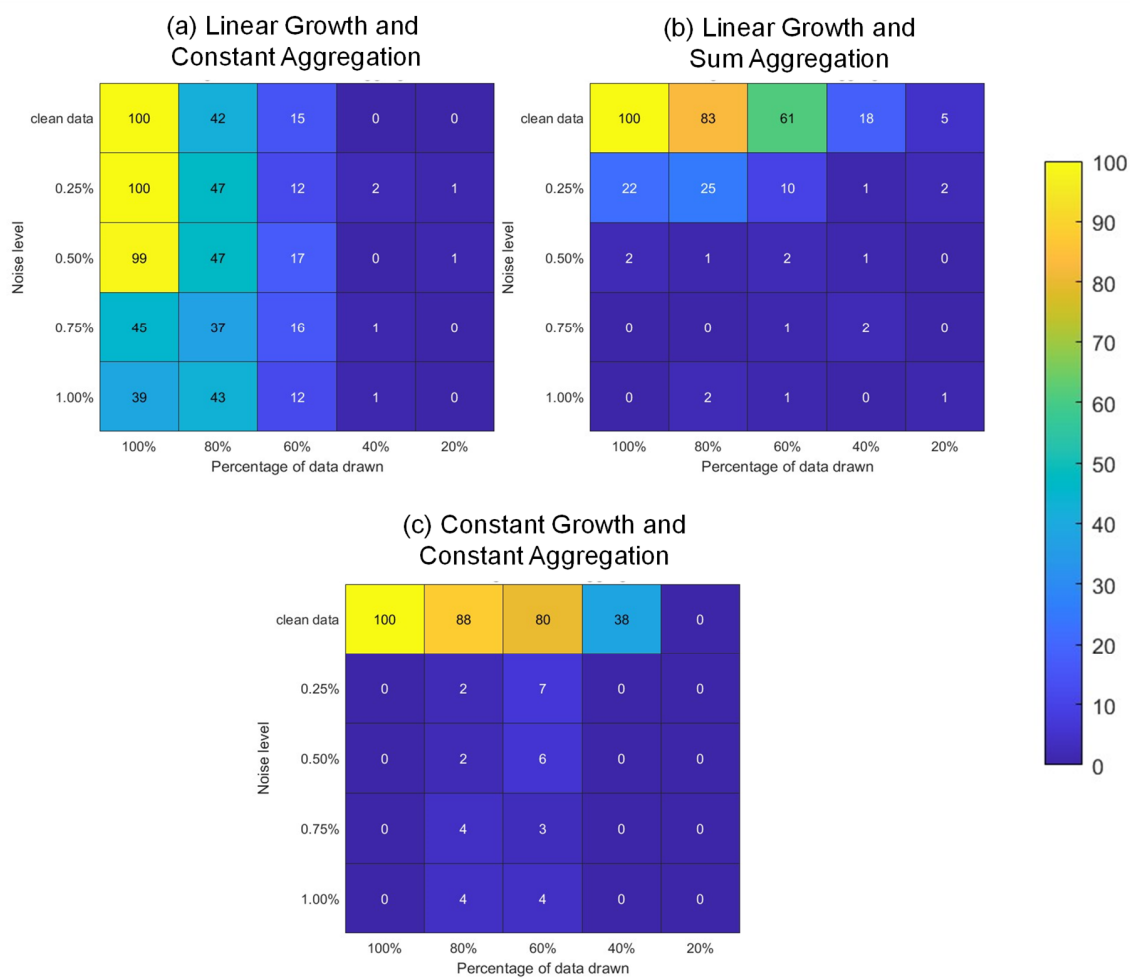

Fig. S5 The success rate (correct identification of PBE terms) of PBE-SPOT on simultaneous growth and aggregation cases: (a) linear growth and constant aggregation, (b) linear growth and sum aggregation and c) constant growth and constant aggregation, across varying noise levels and amount of data. The noise levels tested are 0% (clean data), 0.25%, 0.50%, 0.75% and 1%. The amount of data used are 20%, 40%, 60%, 80% and 100% (data domain as indicated in Section S1.3). For each combination of noise level and data percentage, 100 random samples were tested.

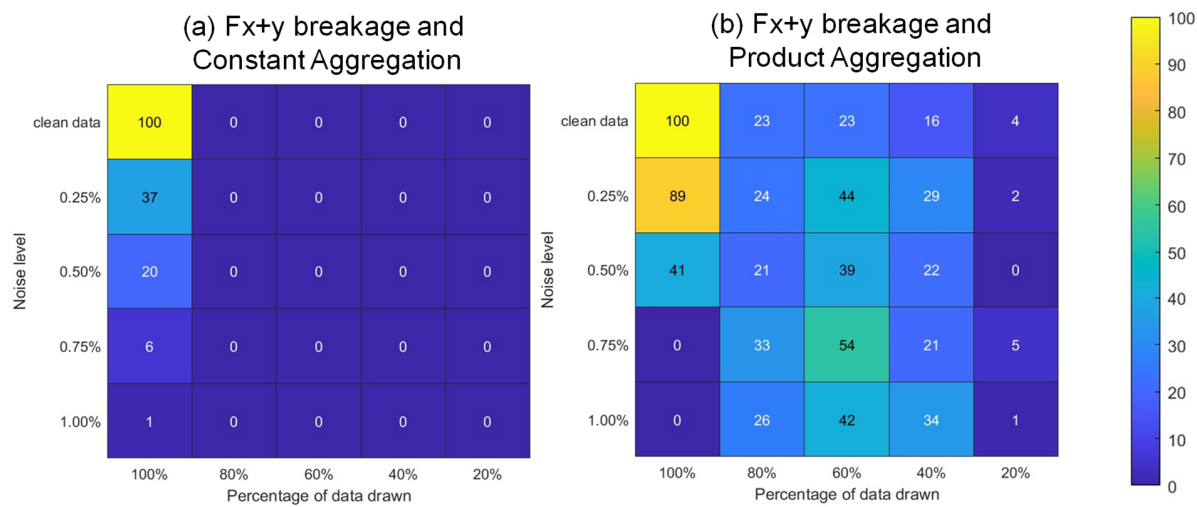

Fig. S6 The success rate (correct identification of PBE terms) of PBE-SPOT on simultaneous Fx+y breakage coupled with various aggregation cases: (a) Fx+y breakage and constant aggregation and (b) Fx+y breakage and product aggregation, across varying noise levels and amount of data. The noise levels tested are 0% (clean data), 0.25%, 0.50%, 0.75% and 1%. The amount of data used are 20%, 40%, 60%, 80% and 100% (data domain as indicated in Section S1.3). For each combination of noise level and data percentage, 100 random samples were tested.

# REFERENCES

1. D. Ramkrishna, *Population balances: Theory and applications to particulate systems in engineering* (Elsevier, 2000).
2. M. V. Smoluchowski, Brownsche Molekularbewegung und Koagulation von Kolloidteilchen. *Phys. Zeits.* **17**, 585-599 (1916).
3. A. J. Griggs, J. J. Stickel, J. J. Lischeske, A mechanistic model for enzymatic saccharification of cellulose using continuous distribution kinetics I: depolymerization by EGI and CBHI. *Biotechnology and bioengineering* **109**, 665-675 (2012).
4. A. J. Griggs, J. J. Stickel, J. J. Lischeske, A mechanistic model for enzymatic saccharification of cellulose using continuous distribution kinetics II: cooperative enzyme action, solution kinetics, and product inhibition. *Biotechnology and Bioengineering* **109**, 676-685 (2012).
5. N. Lebaz, A. Cockx, M. Spérandio, J. Morchain, Population balance approach for the modelling of enzymatic hydrolysis of cellulose. *The Canadian Journal of Chemical Engineering* **93**, 276-284 (2015).
6. F. Ahamed, H.-S. Song, C. W. Ooi, Y. K. Ho, Modelling heterogeneity in cellulose properties predicts the slowdown phenomenon during enzymatic hydrolysis. *Chemical Engineering Science* **206**, 118-133 (2019).
7. F. Ahamed, H. S. Song, Y. K. Ho, Modeling coordinated enzymatic control of saccharification and fermentation by Clostridium thermocellum during consolidated bioprocessing of cellulose. *Biotechnology and bioengineering* **118**, 1898-1912 (2021).
8. Y. K. Ho, P. Doshi, H. K. Yeoh, G. C. Ngoh, Interlinked population balance and cybernetic models for the simultaneous saccharification and fermentation of natural polymers. *Biotechnology and bioengineering* **112**, 2084-2105 (2015).
9. S. Misra, S. Mishra, Charging kinetics of dust in interplanetary space plasma. *Monthly Notices of the Royal Astronomical Society* **432**, 2985-2993 (2013).
10. A. A. Lipovka, Solution of a set of population-balance equations for atomic and molecular quantum levels in some particular cases. *Astronomy Reports* **39**, 347-350 (1995).
11. S. Mishra, T. Sana, Distribution of charge on floating dust particles over sunlit locations on Moon. *Monthly Notices of the Royal Astronomical Society* **508**, 4332-4341 (2021).
12. E. Sherer, R. Hannemann, A. Rundell, D. Ramkrishna, Analysis of resonance chemotherapy in leukemia treatment via multi-staged population balance models. *Journal of theoretical biology* **240**, 648-661 (2006).
13. E. Sherer, R. E. Hannemann, A. Rundell, D. Ramkrishna, Estimation of likely cancer cure using first-and second-order product densities of population balance models. *Annals of biomedical engineering* **35**, 903-915 (2007).
14. P. Y. Huang, J. D. Hellums, Aggregation and disaggregation kinetics of human blood platelets: Part I. Development and validation of a population balance method. *Biophysical journal* **65**, 334-343 (1993).
15. H. Wright, D. Ramkrishna, Solutions of inverse problems in population balances—I. Aggregation kinetics. *Computers & chemical engineering* **16**, 1019-1038 (1992).
16. A. Sathyagal, D. Ramkrishna, G. Narsimhan, Solution of inverse problems in population balances-II. Particle break-up. *Computers & chemical engineering* **19**, 437-451 (1995).

17. I. Nopens, N. Nere, P. A. Vanrolleghem, D. Ramkrishna, Solving the inverse problem for aggregation in activated sludge flocculation using a population balance framework. *Water Science and Technology* **56**, 95-103 (2007).
18. A. Groh, J. Krebs, Improved solution methods for an inverse problem related to a population balance model in chemical engineering. *Inverse problems* **28**, 085006 (2012).
19. N. B. Raikar, S. R. Bhatia, M. F. Malone, M. A. Henson, Self-similar inverse population balance modeling for turbulently prepared batch emulsions: Sensitivity to measurement errors. *Chemical engineering science* **61**, 7421-7435 (2006).
20. H. Wright, R. Muralidhar, T. Tobin, D. Ramkrishna, Inverse problems of aggregation processes. *Journal of statistical physics* **61**, 843-863 (1990).
21. A. W. Mahoney, F. J. Doyle III, D. Ramkrishna, Inverse problems in population balances: Growth and nucleation from dynamic data. *AIChE journal* **48**, 981-990 (2002).
22. J. Chakraborty, J. Kumar, M. Singh, A. Mahoney, D. Ramkrishna, Inverse problems in population balances. Determination of aggregation kernel by weighted residuals. *Industrial & Engineering Chemistry Research* **54**, 10530-10538 (2015).
23. S. L. Brunton, J. N. Kutz, *Data-driven science and engineering: Machine learning, dynamical systems, and control* (Cambridge University Press, 2022).
24. S. L. Brunton, J. L. Proctor, J. N. Kutz, Discovering governing equations from data by sparse identification of nonlinear dynamical systems. *Proceedings of the national academy of sciences* **113**, 3932-3937 (2016).
25. K. Champion, B. Lusch, J. N. Kutz, S. L. Brunton, Data-driven discovery of coordinates and governing equations. *Proceedings of the National Academy of Sciences* **116**, 22445-22451 (2019).
26. K. Kaheman, J. N. Kutz, S. L. Brunton, SINDy-PI: a robust algorithm for parallel implicit sparse identification of nonlinear dynamics. *Proceedings of the Royal Society A* **476**, 20200279 (2020).
27. M. Quade, M. Abel, J. Nathan Kutz, S. L. Brunton, Sparse identification of nonlinear dynamics for rapid model recovery. *Chaos: An Interdisciplinary Journal of Nonlinear Science* **28** (2018).
28. P. Zheng, T. Askham, S. L. Brunton, J. N. Kutz, A. Y. Aravkin, A unified framework for sparse relaxed regularized regression: SR3. *IEEE Access* **7**, 1404-1423 (2018).
29. K. Champion, P. Zheng, A. Y. Aravkin, S. L. Brunton, J. N. Kutz, A unified sparse optimization framework to learn parsimonious physics-informed models from data. *IEEE Access* **8**, 169259-169271 (2020).
30. U. Fasel, J. N. Kutz, B. W. Brunton, S. L. Brunton, Ensemble-SINDy: Robust sparse model discovery in the low-data, high-noise limit, with active learning and control. *Proceedings of the Royal Society A* **478**, 20210904 (2022).
31. D. A. Messenger, D. M. Bortz, Weak SINDy for partial differential equations. *Journal of Computational Physics* **443**, 110525 (2021).
32. D. A. Messenger, D. M. Bortz, Weak SINDy: Galerkin-based data-driven model selection. *Multiscale Modeling & Simulation* **19**, 1474-1497 (2021).
33. D. Bertsimas, W. Gurnee, Learning sparse nonlinear dynamics via mixed-integer optimization. *Nonlinear Dynamics* **111**, 6585-6604 (2023).
34. Z. Chen, Y. Liu, H. Sun, Physics-informed learning of governing equations from scarce data. *Nature communications* **12**, 6136 (2021).

35. S. H. Rudy, S. L. Brunton, J. L. Proctor, J. N. Kutz, Data-driven discovery of partial differential equations. *Science advances* **3**, e1602614 (2017).
36. A. Blumer, A. Ehrenfeucht, D. Haussler, M. K. Warmuth, Occam's razor. *Information processing letters* **24**, 377-380 (1987).
37. W. T. Scott, Analytic studies of cloud droplet coalescence I. *Journal of Atmospheric Sciences* **25**, 54-65 (1968).
38. R. M. Ziff, E. McGrady, The kinetics of cluster fragmentation and depolymerisation. *Journal of Physics A: Mathematical and General* **18**, 3027 (1985).
39. T. Ramabhadran, T. Peterson, J. Seinfeld, Dynamics of aerosol coagulation and condensation. *AIChE Journal* **22**, 840-851 (1976).
40. S. Kumar, D. Ramkrishna, On the solution of population balance equations by discretization—I. A fixed pivot technique. *Chemical Engineering Science* **51**, 1311-1332 (1996).
41. J. Orehek, D. Teslic, B. Likozar, Continuous crystallization processes in pharmaceutical manufacturing: A review. *Organic Process Research & Development* **25**, 16-42 (2020).
42. C. Selomulya, G. Bushell, R. Amal, T. D. Waite, Understanding the role of restructuring in flocculation: The application of a population balance model. *Chemical Engineering Science* **58**, 327-338 (2003).
43. S. M. Iveson, Limitations of one-dimensional population balance models of wet granulation processes. *Powder Technology* **124**, 219-229 (2002).
44. C. Kotoulas, C. Kiparissides, A generalized population balance model for the prediction of particle size distribution in suspension polymerization reactors. *Chemical Engineering Science* **61**, 332-346 (2006).
45. Y. K. Ho, P. Doshi, H. K. Yeoh, Modelling simultaneous chain‐end and random scissions using the fixed pivot technique. *The Canadian Journal of Chemical Engineering* **96**, 800-814 (2018).
46. J. Matt (2020) Extract linearly independent subset of matrix columns. (MATLAB Central File Exchange).